\documentclass[submission, Proceedings]{SciPost}
\usepackage{caption}
\usepackage{subcaption}

\begin{document}

\begin{center}{\Large \textbf{
The impact of ATLAS $V$+jet measurements on PDF fits
}}\end{center}

\begin{center}
E. I. Conroy\textsuperscript{1*}, on behalf of the ATLAS collaboration
\end{center}

\begin{center}
{\bf 1} University of Oxford, Oxford, United Kingdom
\\
* eimear.conroy@physics.ox.ac.uk
\end{center}

\begin{center}
\today
\end{center}


\section*{Abstract}
{\bf
The production of $W$/$Z$-bosons in association with jets is an important test of perturbative QCD predictions and also yields information about the parton distribution functions of the proton. We present fits to determine PDFs using inclusive $W$/$Z$-boson and $W$/$Z$+jets measurements from the ATLAS experiment at the LHC. The ATLAS measurements are used in combination with deep-inelastic scattering data from HERA. An improved determination of the sea-quark densities at high Bjorken-$x$ is seen, while confirming a strange-quark density similar in size to the up- and down-sea-quark densities in the range $x<0.02$ found by previous ATLAS analyses.
}


\section{Introduction}
\let\thefootnote\relax\footnotetext{Copyright 2021 CERN for the benefit of the ATLAS Collaboration. CC-BY-4.0 license.}

\label{sec:intro}
Precise knowledge of Parton Distribution Functions (PDFs) is crucial for both Standard Model and Beyond Standard Model physics at hadron colliders such as the Large Hadron Collider (LHC).
Previous studies \cite{epWZ16} conducted by the ATLAS experiment \cite{ATLAS} at the LHC have demonstrated the power of ATLAS inclusive $W$ and $Z$ boson production data to constrain the flavour of the down-type sea. In this ATLASepWZ16 fit, ATLAS data is combined with HERA deep inelastic scattering (DIS) data \cite{HERA} to extend the range in Bjorken-$x$ and negative squared four-momentum transfer $Q^2$. 

This previous ATLAS fit found the strange quark density, $R_S=(s+\Bar{s})/(\Bar{u}+\Bar{d})$, to be unsupressed at low-$x$, in tension with the indications from fixed-target neutrino-induced DIS experiments. The $(\Bar{d}-\Bar{u})$ distribution, a topic of debate over recent decades, was also found to exhibit tension with measurements by the fixed-target Drell-Yan experiment E866 \cite{E866}, which found the distribution to be positive at high-$x$ while the ATLASepWZ16 fit preferred a slightly negative value, though compatible with zero within uncertainties.

The addition of vector boson in association with jets ($V$+jets, where $V=W,Z$) production data would allow a fit to be sensitive to the gluon at leading-order in quantum chromodynamics (QCD) and to further extend the $x$ and $Q^2$ reach compared to fits including only inclusive vector boson data. This article describes such a fit \cite{Vjets}, resulting in the ATLASepWZVjet20 PDF set. 

\section{Description of Datasets}
HERA combined neutral current (NC) and charged current (CC) DIS data are fit simultaneously with several ATLAS cross-section measurements -- inclusive $W,Z$ production at 7 TeV \cite{epWZ16}, $W$+jet production at 8 TeV \cite{Wjets} and $Z$+jet production at 8 TeV \cite{Zjets}. A cut of $Q^2_\text{min}$ was applied to the DIS data to avoid regions requiring complex treatment. Theoretical predictions are available at next-to-next-to leading-order (NNLO) in QCD for the DIS processes \cite{qcdnum}.The ATLAS cross-section measurements use predictions at next-to leading-order (NLO) in QCD, evolved up to NNLO through the use of $k$-factors, alongside leading-order (LO) electroweak corrections evolved to NLO also using $k$-factors. The details of these predictions are discussed in their respective ATLAS publications \cite{epWZ16}\cite{Wjets}\cite{Zjets}.

The 8 TeV $V$+jets cross-sections used in the fit are the single differential $\text{d}\sigma/\text{d}p_\text{T}^W$ distribution for the $W$+jets data and the double differential $\text{d}\sigma/\text{d}p_\text{T}^\text{jet}\text{d}|y^\text{jet}|$ distribution for the $Z$+jets data. The inclusion of the $V$+jets data in the fit resulted in a better description of the $W$+jets data at high transverse momentum ($p_\text{T}$) and a change to the normalisation of the $Z$+jets spectrum.





\section{Fit Procedure}
\label{sec:another}
A QCD analysis was performed at NNLO using the \textsc{xFitter} framework \cite{xfitter} combined with MINUIT \cite{minuit}, with cross-validation performed using an independent framework. A starting scale of $Q^2_0=1.9$ GeV$^2$ was chosen for DGLAP evolution as it is below the scale of the charm-quark mass.

The distributions of the valence quarks $xu_v$ and $xd_v$, the light sea anti-quarks $x\Bar{u}$, $x\Bar{d}$ and $x\Bar{s}$ and the gluon $xg$ were parameterised at the starting scales using the same framework as HERAPDF2.0 \cite{HERA} and ATLASepWZ16 \cite{epWZ16}. This takes the general form 

\begin{equation*}
    xq_i(x)=A_ix^{B_i}(1-x)^{C_i}P_i(x),
\end{equation*}

\noindent where terms in the polynomial $P_i(x)=(1+D_ix+E_ix^2)e^{F_ix}$ are included only if required by the fit. Additionally, the gluon distribution has an extra negative term for further flexibility.


Requirements, such as the number sum rule, momentum sum rule, $x\Bar{u}=x\Bar{d}$ as $x\xrightarrow{}0$ and the suppression of negative gluon contributions at high-$x$ provide constraints on the fit. In total, 16 free parameters are used in the central fit. 



In addition to experimental uncertainties associated with the input datasets, uncertainties arising from theoretical and parameterisation choices are also considered.  Parameterisation uncertainties are estimated by relaxing constraints or adding extra $D$, $E$, and $F$ parameters, which control the low-$x$ sea, and repeating the fit. Similarly, model uncertainties are estimated by varying theoretical assumptions, such as $Q^2_\text{min}$, $Q^2_0$ and heavy quark masses.



\section{Results}
\begin{figure}[h]
    \centering
    \begin{subfigure}[b]{0.45\textwidth}
        \centering
        \includegraphics[width=\textwidth]{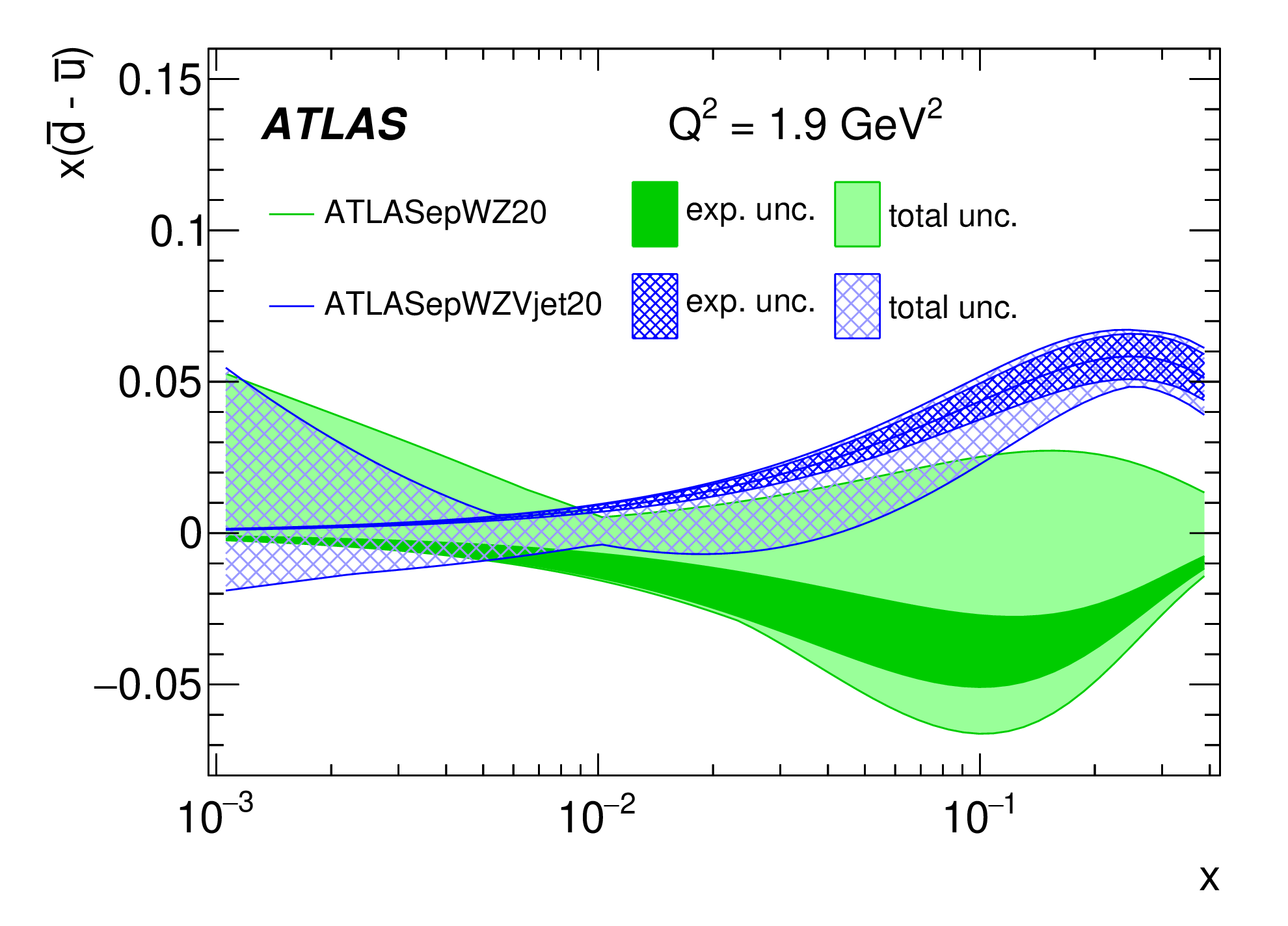}
        \caption{}
        \label{fig:dbarubar}
    \end{subfigure}
    \begin{subfigure}[b]{0.45\textwidth}
        \centering
        \includegraphics[width=\textwidth]{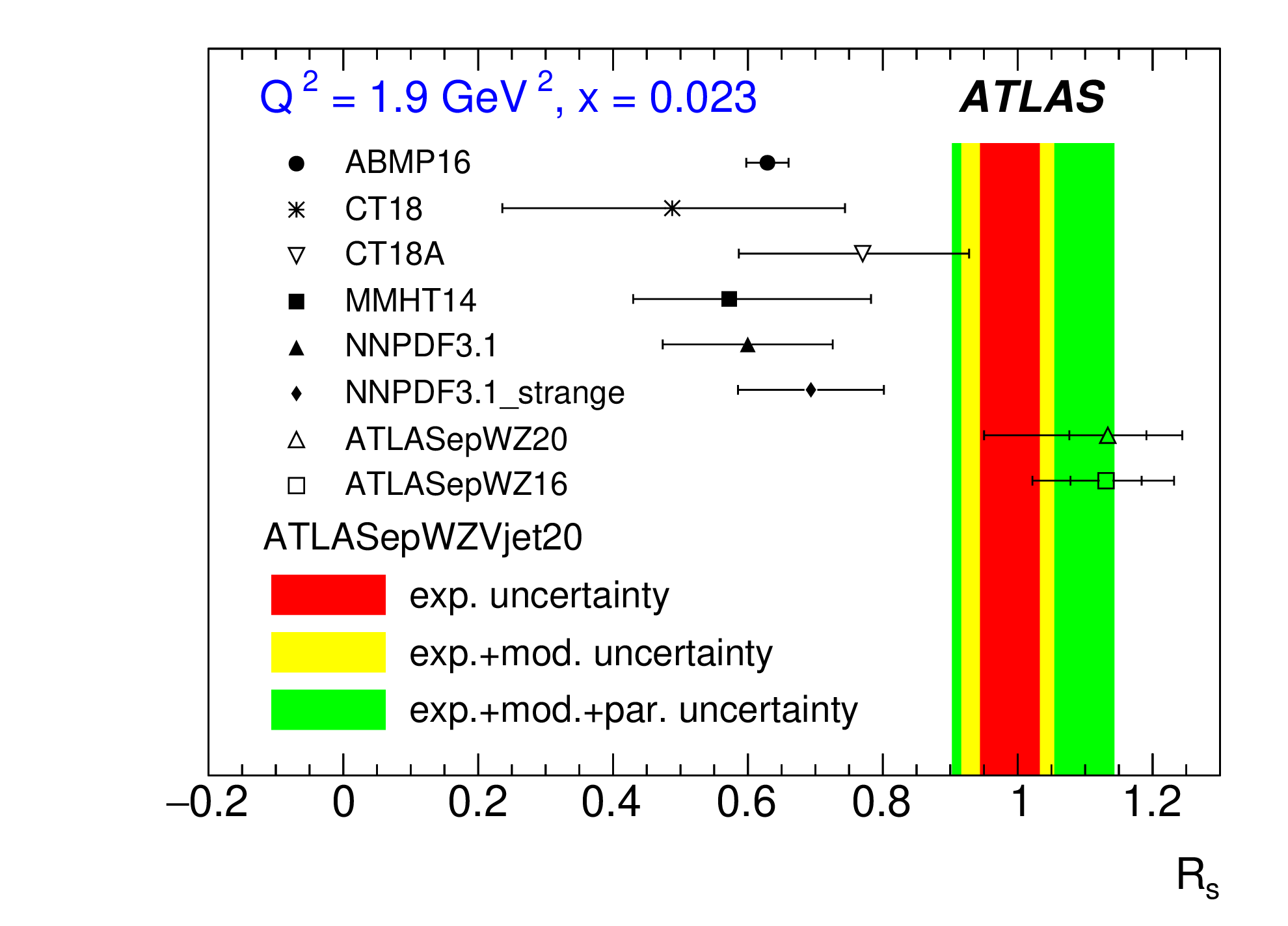}
        \caption{}
        \label{fig:rs}
    \end{subfigure}
    \caption{(a) The $x(\Bar{d}-\Bar{u})$ distribution evaluated at the starting scale $Q^2 = 1.9$ GeV$^2$ as a function of Bjorken-$x$, extracted from ATLASepWZVjet20 \cite{Vjets} (blue) and ATLASepWZ20 (green) - an identical fit with $V$+jets data excluded. (b) $R_s$ evaluated at $x=0.023$ and $Q^2=1.9$ GeV$^2$, with ATLASepWZVjet20 \cite{Vjets} compared to other PDF sets. The ATLASepWZVjet20 experimental, model and parameterisation uncertainty bands are plotted separately.}
\end{figure}

The resulting PDF set, ATLASepWZVjet20, is compared to a fit performed under identical conditions with $V$+jets data excluded, ATLASepWZ20. This comparison showed the inclusion of the $V$+jets data in the fit increases the high-$x$ $x\Bar{d}$ distribution and decreases the high-$x$ $x\Bar{s}$ distribution, with the sum of the two constrained by HERA data. 

This increased high-$x$ $x\Bar{d}$ also affects the $x(\Bar{d}-\Bar{u})$ distribution as shown in Figure \ref{fig:dbarubar}. Without the $V$+jets data, this distribution is slightly negative but compatible with 0 within uncertainties at high-$x$, similar to ATLASepWZ16. With the inclusion of the $V$+jets data, the distribution becomes positive at high-$x$, in better agreement with E866 results up to $x\sim0.1$.

Thorough investigation using a $\chi^2$ scan of the $C_{\Bar{d}}$ parameter found that it is the ATLAS data which prefer this larger high-$x$ $x\Bar{d}$, while HERA data prefer a smaller value. Thus, the inclusion the ATLAS $V$+jets dataset decreases the impact of the HERA data and eliminates a double minimum observed at both high and low values of $C_{\Bar{d}}$ when this data is not included.


The impact of the $V$+jets data on the $R_s$ distribution was also investigated, as shown in Figure \ref{fig:rs}. It was found that ATLASepWZVjet20 retains an unsupressed $R_s$ at low-$x$, in tension with results from global PDF fitters, however, this tension is decreased somewhat with respect to the findings of previous ATLAS fits. The tension is least with the CT18A \cite{ct18a} fit, which includes ATLAS inclusive $W,Z$ data at 7 TeV.

\section{Conclusion}
The impact of vector boson in association with jets production data obtained at 8 TeV by the ATLAS experiment at the LHC on the Parton Distribution Functions of the proton was assessed. These data were fit simultaneously with HERA DIS and ATLAS $W,Z$ inclusive cross-section data resulting in the ATLASepWZVjet20 PDF set. An increased high-$x$ $x\Bar{d}$ distribution was found to bring the $x(\Bar{d}-\Bar{u})$ in the same region into better agreement with fits using E866 data and into tension with previous ATLAS fits. Additionally, the strange-quark density at low-$x$ was found to be unsuppressed, in agreement with previous ATLAS fits, however the previously observed tension with other global fits is reduced.

\end{document}